\documentclass[aps,prstab,reprint,groupedaddress]{revtex4-1}
\usepackage{graphicx}
\usepackage{dcolumn}
\usepackage{bm}
\usepackage{amsmath}
\usepackage{amssymb}
\usepackage{booktabs}
\usepackage{subcaption}
\usepackage{siunitx}
\usepackage{hyperref}
\usepackage[capitalize]{cleveref}

\newcommand{\sgn}{\mathop{\mathrm{sgn}}}
\DeclareMathOperator\Ai{\mathrm{Ai}}
\newcommand\hdot{\ensuremath{\dot{h}}}
\newcommand\asymm{\ensuremath{\mathcal{A}}}
\begin{document}


\title{Estimate of Multi-Shot Laser-Induced Polarization for High Energy Electrons}

\thanks{Work supported by URA., Inc., under contract DE-AC02-76CH03000 with the U.S. Dept. of Energy.}
\author{Katherine D. Ranjbar\thanks{ranjbark24@gmail.com}, Emily Snyder, Alice Snyder}
\affiliation{Earl L. Vandermeulen High School,
   350 Old Post Road, Port Jefferson, NY 11777}
\author{V. H.  Ranjbar }

\affiliation{Brookhaven National Lab, Upton NY 11973}


\date{\today}

\begin{abstract}
The use of an intense ultrashort laser pulse to induce electron polarization has been proposed in existing literature.  The Python programming language is used to recreate the local constant crossed-field approximation (LCFA) with the aim of determining values for transverse polarization given a nonzero initial polarization. It has been shown that over multiple laser shots, lower values of the quantum efficiency parameter are associated with higher transverse polarization output, yet require a greater number of shots to attain maximal polarization. Moreover, the quantum efficiency parameter has been redefined as a function of intensity for a Ti:sapphire laser necessary to induce polarization in the Electron-Ion Collider.
\end{abstract}


\maketitle

\section{Introduction}
The polarization of high energy electrons in a storage ring is an active area of interest for accelerators. For example, several planned storage rings will require polarized electrons.
These include the Electron Ion Collider (EIC), China's proposed Circular Electron Positron Collider (CEPC)~\cite{Duan:2023lyp} as well as for energy calibration in CERN's proposed Future Circular Collider (FCC)~\cite{Blondel:2021zix}. Considering the EIC, it has been designed to investigate the strong force by colliding intense
beams of polarized electrons with protons and nuclei to reveal the arrangement of quarks and gluons
within the atomic nucleus. Polarization of the electron beam is essential for the EIC, and the current plan is to transport highly polarized electrons from a polarized source through a linear accelerator, and then accelerate them to the final energy in a rapid cycling synchrotron using a novel design that minimizes the polarization lost to spin resonances during the acceleration process~\cite{Ranjbar:2018lpo, Skaritka:2018oxj}. It is also known that electrons
circulating in a storage ring will self-polarize due to the Sokolov-Ternov (ST)
effect, a type of radiative spin-polarization whereby a particle’s change in momentum caused by motion
along a bent path results in the ejection of a photon (i.e., synchrotron radiation) and subsequent spin
flip. Since transition to the spin-down state is more likely, up to 92.4\% antiparallel transverse-plane
polarization may be achieved over the course of minutes or hours --- a relatively slow process~\cite{Sokolov:1963zha}.
Both the ST polarization time and the need for multiple bunches in a single store polarized both antiparallel and parallel relative to the dipole guide field make the use of ST polarization impractical.

However, it has been shown that an intense laser pulse can be used to polarize electrons in
a matter of femtoseconds~\cite{DelSorbo:2017fod}. The interaction between an unpolarized beam and an
intense laser field results in the transfer of energy to an outgoing gamma-ray photon through nonlinear
Compton scattering. As in the ST effect, polarization is obtained through the loss of a photon and
resultant spin flip. Experimental results as well as theoretical work have confirmed that the degree of
polarization due to nonlinear Compton scattering varies with laser parameters as well as initial
polarization~\cite{Seipt:2017ckc}.
Laser-induced polarization for high-energy polarized machines like the EIC is worth considering for several reasons: not only is polarization attained on a comparatively shorter time scale, but the necessary parallel polarization (relative to the dipole guide field), which would otherwise be countered by the ST effect, can also be achieved.

The magnitude of polarization~(0.5\%) reported in these studies~\cite{Seipt:2018adi} is much smaller than the~70\% average polarization expected for the EIC program. However we were curious about the repeated application of a laser pulse to the circulating high energy beam and the possibility of building up sizable average polarization over many shots. Initial estimates of the possible energy kick of an \SI{18}{\GeV} electron beam show that even with a laser power on the order
of a gigawatt the energy kick would be negligible. In this paper we deploy the method developed by~\cite{Seipt:2018adi} to study the viability polarizing high energy electron beams using a high-powered laser in a machine like the EIC. 

\begin{figure}[!htb]
  \begin{center}
    \includegraphics[width=\columnwidth]{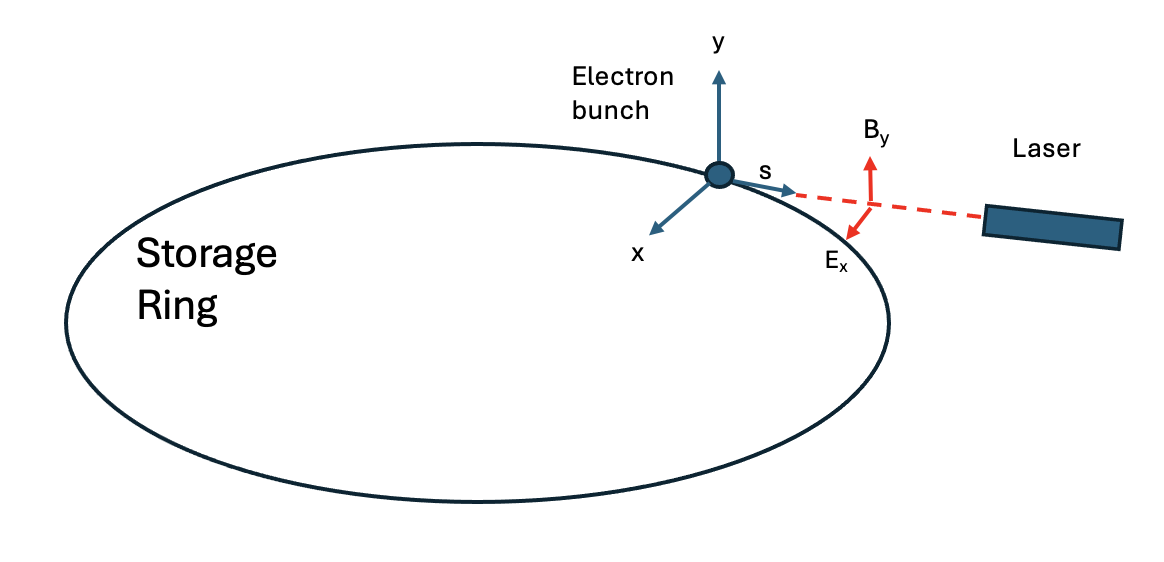}
  \end{center}
   \caption{Layout with a linearly polarized laser incident on an electron bunch stored in an accelerator at \SI{18}{\GeV}. }
    \label{fig1}
\end{figure}

Two methods have been previously proposed for calculating the degree of polarization in final
state electrons. The first uses a density matrix to calculate the quantum state of the electrons. The alternative method, known as
the local constant crossed-field approximation (LCFA), estimates the results of the density matrix
for a linearly polarized laser. The results of the LCFA have been shown to be consistent with those of
the density matrix~\cite{Seipt:2018adi}.
According to the LCFA approximation, the electron polarization vector, 
describing the degree of polarization within a system across the transverse plane, can be represented by:
\begin{widetext}
\begin{equation}
  \Xi_{y'} = \frac
    {\int d{\tau}dt \left[
      \Xi_{y}\Ai_{1}(z(t,\tau))+\Xi_{y}\frac{2\Ai'(z(t,\tau))}{z(t,\tau)} +
      \frac{t}{1-t}\frac{\Ai(z(t,\tau))}{\sqrt{z(t,\tau)}}\sgn(\dot{h}(\tau))
    \right]}
    {\int d{\tau}dt \left[
      \Ai_{1}(z(t,\tau)) + \left(1+\frac{t^2}{2(1-t)}\right)\frac{2\Ai'(z(t,\tau))}{z(t,\tau)} +
      \Xi_{y}\sgn(\dot{h}(\tau))t\frac{\Ai(z(t,\tau))}{\sqrt{z(t,\tau)}}
    \right]}.
\label{eq1}
\end{equation}
\end{widetext}
Here, $\Xi_{y}$ describes the initial polarization of the beam relative to the vertical plane in a coordinate system with the storage ring and laser as shown in~\cref{fig1}.
The integration variables $t$ and $\tau$
are the fractional momentum transfer and proper time, respectively.
The argument $z$ of the Airy functions $\Ai(z)$, their derivatives
$\Ai'(z)$, and their integrals $\Ai_1(z) = \int_z^{\infty}dx \Ai(x)$
is given by
$z=z(t,\tau)=\left [\frac{t}{1-t}\frac{1}{| \dot{h}(\tau)|\chi_e}\right ]^{\frac{2}{3}}$, where
the pulse shape function $h(\phi) = \cos(\phi+\phi_{\mathrm{CE}})\cos^2(\frac{\pi \phi}{2 \Delta \phi}) \Theta(\Delta \phi - |\phi|)$ for the
vector potential, and so its time derivative $\dot{h}$ describes the shape of the electric field.
The shape and asymmetry of a magnetic field produced by a laser are limited due to the inability to
produce ``unipolar'' fields (where $h(-\infty) \ne h(\infty)$).
In the semiclassical limit, the polarization is proportional to the
asymmetry $\asymm$ of $\hdot$:
\begin{equation}
  \Xi_{y'} =
    -\frac{3\chi_e}{10} \frac{\int d\tau\, |\hdot(\tau)|\hdot(\tau)}{\int d\tau\, |\hdot(\tau)|} =
    \frac{3\chi_e}{10}\asymm.
\end{equation}
The asymmetry is maximal at $|\asymm| \approx 0.17$ when the carrier
envelope phase $\phi_{\mathrm{CE}}=\pi/2$ and the pulse duration
$\Delta\phi = 3.27$, and effectively becomes zero for $\Delta\phi > 8$~\cite{Seipt:2018adi}.

\section{Methodology}
\begin{figure}[!htb]
  \begin{center}
    \large \textbf{Transverse polarization $\Xi_{y}$ vs.~quantum efficiency parameter $\chi_e$}
    \includegraphics[width=\columnwidth]{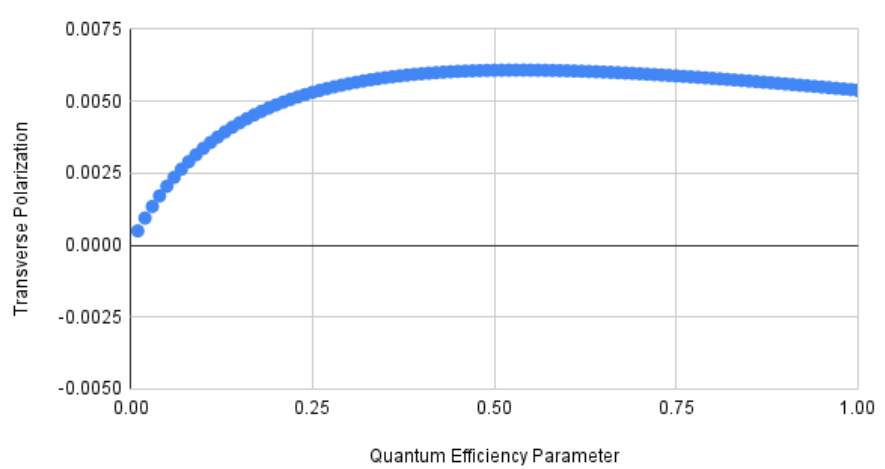}
  \end{center}
   \caption{Transverse polarization~$\Xi_{y}$ as a function of quantum efficiency parameter~$\chi_e$. The results are consistent with those presented in~\cite{Seipt:2018adi}, with~$\chi_e$ peaking at~0.5.}
    \label{xpol}
\end{figure}

A Python program~\cite{LaserSim:2023} was used to find final polarization amplitudes given
input values of $\Delta\phi$, $\phi_{\mathrm{CE}}$, and~$\chi_e$.
The LCFA approximation, \cref{eq1}, was used to create a function
taking input values of the quantum efficiency parameter~$\chi_e$ and returning values of transverse
polarization~$\Xi_{y}$. This was accomplished by splitting~\cref{eq1} into four different integrals, and then
integrating them separately. Since the integrands in question diverge as $\chi_e$ approaches zero, it was necessary to perform a series
of substitutions. In order to confirm the validity of the code, values for transverse polarization~$\Xi_{y}$ were plotted as
a function of the quantum efficiency parameter~$\chi_e$ for zero initial polarization. The results were consistent with those presented in existing literature~\cite{Seipt:2018adi}, with a peak of~0.006 occurring
at $\chi_e = 0.5$ (\cref{xpol}).

The quantum efficiency parameter, $\chi_e$, is roughly defined as a measure of field strength. In
traditional radiative polarization, where the electromagnetic field is weak, $\chi_e = e\sqrt{|(F^{\mu\nu} p_{\nu})^2|}/m^3 \ll 1$~\cite{Baier}, where $e$ is the fundamental electric charge, $F$ is the electromagnetic tensor,
$p$ is momentum, and $m$ is the electron mass. $\chi_e$ can also be defined as the product of the nonlinearity
parameter 
$\xi = 8.55 \sqrt{\lambda^2[\mathrm{\mu m}]I\left[10^{20}{{\mbox{W}}\over{\mbox{cm}^2}}\right]} $, with wavelength $\lambda$ and intensity $I$, and the laser
frequency in the electron rest frame $ b = {{k \cdot p}\over{m^2}}$ , with $k \cdot p$ being the dot product of the photon
momentum vector  $k$ and the electron momentum vector  $p$. In a strong field where rapid polarization
occurs due to frequent spin flips, $\chi_e = \xi b$ is approximately~1, and $\xi \gg 1$~\cite{Seipt:2018adi}. Since
the electron momentum vector that constitutes $b$ is defined as $p = m \gamma v$, where $v$ represents the
electron velocity and $\gamma$ is the Lorentz factor, it stands that $b = {{k \cdot (m \gamma v)}\over{m^2}} = {{k \cdot (\gamma v)}\over{m}}$ (note that Planck’s constant and the speed of light are not shown, as they are held equal to 1 in order to produce a
measurement in terms of energy). The frequency in the electron rest frame without respect to mass, $k \cdot (\gamma v) = \omega_L$,
is equal to $2 \gamma \omega_0$, where $\omega_0$ is the frequency in the lab rest frame~\cite{Heinzl:2022lly}. This is 
because $\omega_0 = 2 \pi f_0$, in which $f_0$, the linear frequency, can be rewritten as $c/\lambda_0$ to yield $\omega_0 = {{2 \pi c}/{\lambda_0}}$.
In the case of the electron rest frame, $\omega_L={{2\pi c}/{\lambda_L}} = {{2 \pi c (2 \gamma)}/{\lambda_0}}$, as $\lambda_L = {{\lambda_0}\over{\gamma(1-\cos(\phi))}}$ due to
the Doppler effect, and it is assumed that the laser hits the beam head-on from an angle of $\pi$.  $\omega_0$ can
then be substituted into the equation, giving $\omega_L= 2 \gamma\omega_0$. Therefore,
\begin{equation}
    \chi_e=8.55\sqrt{\lambda^2[\mathrm{\mu m}]I\left[10^{20}{{\mbox{W}}\over{\mbox{cm}^2}}\right]}\left({{2\gamma\omega_o}\over{m}}\right).
    \label{eq2}
\end{equation}

\section{{Results}}
\begin{figure*}[!tbh]
    \large \textbf{ Number of Shots vs. Transverse Polarization $\Xi_{y}$ }
  \begin{center}
  \subfloat[$\chi_e = 0.01$]
  {\includegraphics[width=0.49\textwidth]{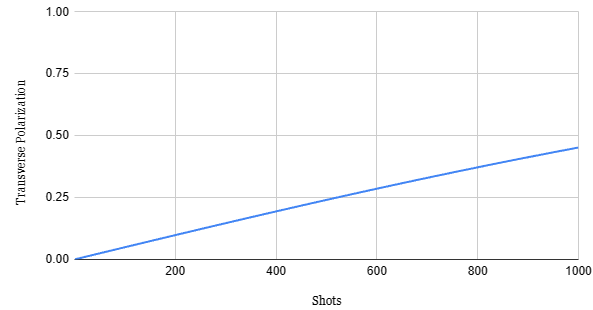}\label{fig2a}}
  \subfloat[$\chi_e = 0.1$]
  {\includegraphics[width=0.49\textwidth]{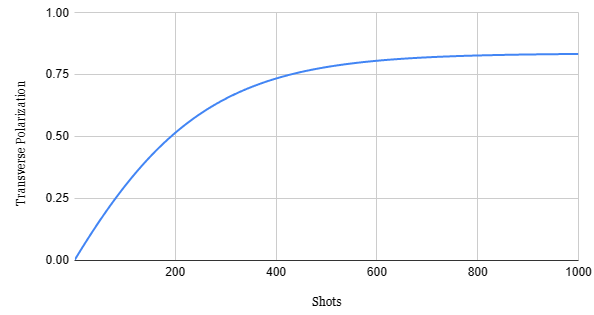}\label{fig2b}}\\
  \subfloat[$\chi_e = 0.5$]
  {\includegraphics[width=0.49\textwidth]{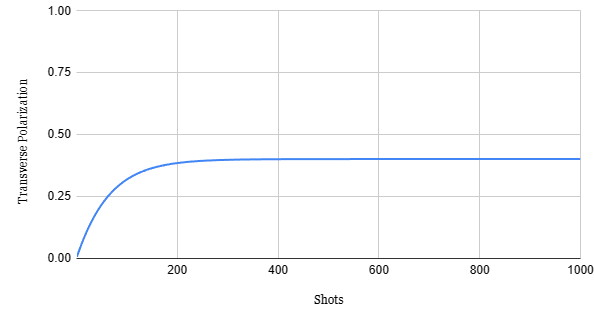}\label{fig2c}}
  \subfloat[$\chi_e = 0.7$]
  {\includegraphics[width=0.49\textwidth]{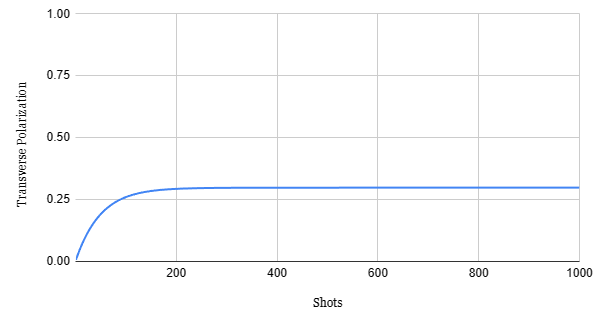}\label{fig2d}}
    \end{center}
   \caption{ Number of shots vs.~transverse polarization for $\chi_e$ = 0.01 (a), 0.1 (b), 0.5 (c), and 0.7 (d). Although polarization plateaus at greater values when $\chi_e$ is low, the number of shots (and by extension time) necessary to achieve maximal polarization is longer. }
   \label{fig2}
  
\end{figure*}

To observe the effect of multiple laser shots, 
the calculated transverse polarization was reused as the input
1000 times given an initial polarization of~0 and
$\chi_e$~values of 0.01~(\cref{fig2a}), 0.1~(\cref{fig2b}), 0.5~(\cref{fig2c}), and 0.7~(\cref{fig2d}).
For $\chi_e= 0.5$, the polarization plateaus at about 40\% after 350 shots.
In the case of the EIC storage ring, with a revolution frequency of about~\SI{12}{\us} or \SI{83}{\kHz}, advanced  Ti:Sapphire laser systems should be able to match if not exceed that described in~\cite{keller2003ultrafast}. In this case the maximal polarization can be achieved in about \SI{4.2}{\ms}. Moreover, for $\chi_e= 0.7$,
polarization levels are at 29.7\% after 250 shots, and for $\chi_e = 0.1$,
83.3\% after
800 shots or \SI{9.6}{\ms}. By contrast, when $\chi_e = 0.01$, polarization increases
linearly at approximately 0.045\% per shot (within the domain of 0 to 1000 shots). As shown in \cref{fig2},
lower quantum efficiency values correspond to greater maximal polarization. However, the number of
shots necessary to reach the plateau threshold --- and to an extent time taken --- similarly increases. In the
case of the EIC, rapid polarization is preferable, particularly for parallel-polarized bunches that, if
exposed to the electromagnetic field for a prolonged period of time, will experience depolarization due
to the ST effect. In order to optimize duration and final state polarization, a $\chi_e$ value around 0.1 is ideal.

\begin{figure}
   \centering
   \large \textbf{Laser power vs.~quantum efficiency parameter $\chi_e$}
   \includegraphics[width=\columnwidth]{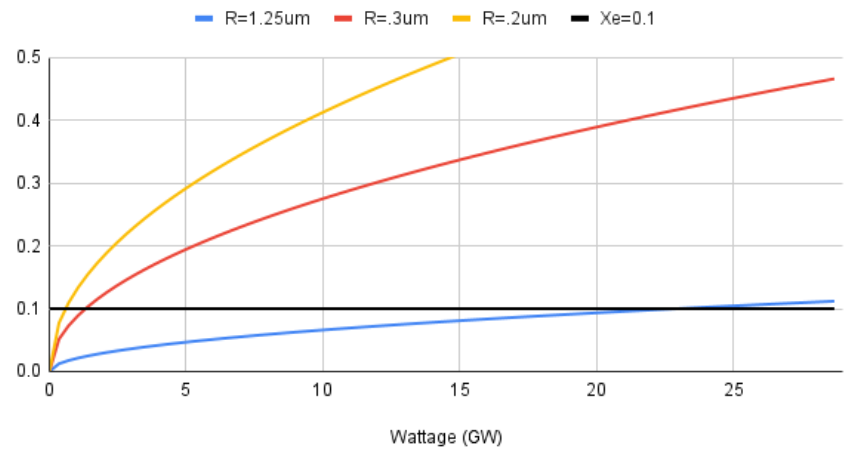}
   \caption{Plot of laser power versus the quantum efficiency parameter~$\chi_e$,
     for beam spot size radii of \SI{1.25}{\um}~(blue), \SI{0.3}{\um}~(red), and
     \SI{0.2}{\um}~(yellow).}
   \label{fig3}
\end{figure}

Using \cref{eq2}, the quantum efficiency parameter~$\chi_e$
can be represented as a function of
laser power. For a typical Ti:sapphire laser, the value of $\omega_0$ is approximately \SI{1.55}{\eV}. When substituted into
$b = 2 \gamma \omega_0/m$, along with a $\gamma$ value of about 35300 for the EIC and electron mass of \SI{0.511}{\MeV}, the frequency in the electron rest frame was calculated to be approximately 0.214. Thus, when $b$ was
substituted into \cref{eq2}, along with a value of $\lambda=\SI{800}{\nm}$ for a standard Ti:sapphire laser, the equation
could be rewritten as $\chi_e = 1.46 \sqrt{I\left[10^{20}{{\mbox{W}}\over{\mbox{cm}^2}}\right]}$, or $\chi_e = 1.46 \sqrt{{{\left[\mbox{W}\right]}\over{(\pi r^2 \times 10^{20})}}}$ when
intensity is normalized with respect to spot size (i.e., the laser beam’s minimum diameter after passing through a focusing lens, given by the formula $S = \frac{4M^2 \lambda f}{\pi d}$ with beam quality parameter $M^2$, wavelength $\lambda$, lens focal length $f$, and diameter at the lens surface $d$).
The laser power for hypothetical fixed spot sizes of radii \SI{1.25}{\um}, \SI{0.3}{\um}, and \SI{0.2}{\um} is plotted against~$\chi_e$, in~\cref{fig3}.

It can be seen that \cref{eq2}, with~$\chi_e$ set equal to~0.1, is satisfied by a power of roughly \SI{23}{\GW} for
a spot size radius of \SI{1.25}{\um}, \SI{1.3}{\GW} for a spot size radius of \SI{0.3}{\um},
and \SI{0.59}{\GW} for a spot size radius of \SI{0.2}{\um}.

\section{{Discussion}}
The exact explanation as to why lower values of the quantum efficiency parameter~$\chi_e$ correlate
with greater final state polarization remains somewhat ambiguous. Evidently, repeated interaction with
the laser pulse results in non-trivial effects that contradict expectations based on the behavior of~$\chi_e$
when initial polarization equals zero. Perhaps the same mechanism responsible for polarization decline
as illustrated in \cref{fig2} when quantum efficiency values exceed~0.5 is at work. Likely, this mechanism
relates to the envelope asymmetry of the ultrashort laser pulse, whereby polarization builds up in the
first half cycle of the monochromatic wave before being partially negated, resulting in an incomplete
cancellation. When initial polarization equals zero ($\Xi_{y} = 0$),
the first two terms of \cref{eq1} vanish;
once incorporated for a nonzero initial polarization,
the proportional relationship between the Airy functions and $\Xi_{y}$ that constitutes these
terms alters the behavior of the function in response to various $\chi_e$ values.

Assuming the validity of the correlation observed in \cref{fig2}, lower quantum efficiency values
are preferable, as they are associated with weaker field strengths and higher asymptotic polarization values.  Typically, the weaker a laser’s field strength,
the higher its potential repetition rate. Thus, a greater number of shots can be fired per second, increasing the
frequency of interaction with the beam.
In our case a laser with power on the order of a gigawatt should be able to match the EIC revolution frequency~\cite{keller2003ultrafast}, and thus a higher laser repetition rate could be used to polarize longer bunches. For example, in the case of the EIC the expected RMS bunch length is on the order of \SI{30}{\ps}.
In the example calculated, the laser envelope pulse length is on the order of \SI{2}{\fs}, which would require 15000 shots to cover the whole RMS length. In this case it would take 144~seconds or 2.4~minutes to polarize a single \SI{30}{\ps} bunch up to the 83\% level using a laser operating at \SI{83}{\kHz}. This is within the bunch lifetime/replacement time for the proposed EIC electron storage ring (ESR), about~6 minutes~\cite{Skaritka:2018oxj}. However the EIC will store 290~bunches at \SI{18}{\GeV}, which would require a laser capable of achieving about \SI{30}{\MHz}, or else  multiple lasers would need to be used.

\section{Conclusions}
The use of an intense ultrashort laser pulse to induce electron polarization has been previously proposed~\cite{Seipt:2018adi,Seipt:2017ckc}. The Python
programming language is used to replicate the local constant crossed-field approximation with the aim of determining values for transverse polarization given a nonzero initial
polarization. It has been shown that polarization for lower values of the quantum efficiency parameter, $\chi_e$ (i.e., weaker field strengths) yield higher final transverse polarization output, yet require a larger number of shots to achieve maximal polarization. The mechanisms responsible for this phenomenon are worth further investigation, particularly those relating to envelope asymmetry and the associated cancellations. While a greater number of shots is necessary to maximize polarization when $\chi_e$ is small, the repetition rate for weaker field strengths is typically greater, potentially allowing more electrons to be polarized. 

Thus, assuming a spot size typical of a Ti:sapphire laser, a laser power corresponding to $\chi_e$ values falling between 0.01 and 0.1 is optimal. In the case of the planned EIC, the revolution frequency of \SI{83}{\kHz} with a \SI{0.6}{\GW} laser matching this pulse rate could polarize a \SI{2}{\fs} slice of an electron bunch in \SI{9.6}{\ms} or a \SI{30}{\ps} bunch in 144~seconds. This could provide a means of countering polarization loss due the ST effect for parallel polarized bunches and spin diffusion effects. Moreover, depolarization brought on by resonances and lattice imperfections in the machine can be avoided or corrected for by polarizing the beam just prior to collision, lessening reliance on the magnetic manipulation of bunch spins. Beyond the EIC, laser-induced polarization also has potential applications in CERN’s Future Circular Collider and in China's proposed CEPC machine.

\section*{Acknowledgments}

The authors would like to thank S.~Snyder for useful discussions and assistance with calculating the numeric integrals.  Work supported by URA., Inc., under contract DE-AC02-76CH03000 with the U.S. Dept. of Energy.

\bibliography{LaserPol}

\end{document}